\documentclass[aps,reprint,amsmath,amssymb,superscriptaddress,showpacs,prb]{revtex4-2}
\usepackage{amsmath}  
\usepackage{amsfonts} 
\usepackage{graphicx} 
\usepackage{color}
\usepackage{lipsum}
\usepackage{dcolumn}
\usepackage{bm}
\usepackage{enumitem}

\newcommand{\vecp}{\bm{p}}
\newcommand{\veck}{\bm{k}}
\newcommand{\vece}{\bm{e}}
\newcommand{\vecv}{\bm{v}}
\newcommand{\vecu}{\bm{u}}
\newcommand{\vecr}{\bm{r}}
\newcommand{\vecR}{\bm{R}}
\newcommand{\vecE}{\bm{E}}

\newcommand{\veczero}{\bm{0}}

\begin{document}

\title{A pedestrian approach to Einstein's formula \boldmath{$E=mc^2$}
  with an application to photon dynamics}

\author{A.~V.~Nenashev}
\affiliation{Department of Physics and Material Sciences Center,
Philipps-University, D-35032 Marburg, Germany}
\affiliation{Institute of Semiconductor Physics, 630090 Novosibirsk, Russia}
\affiliation{Novosibirsk State University, 630090 Novosibirsk, Russia}

\author {S.~D.~Baranovskii}
\affiliation{Department of Physics and Material Sciences Center,
Philipps-University, D-35032 Marburg, Germany}
\affiliation{Department f\"{u}r Chemie, Universit\"{a}t zu K\"{o}ln, Luxemburger Strasse 116, 50939 K\"{o}ln, Germany}

\author{F.~Gebhard}
\affiliation{Department of Physics and Material Sciences Center,
Philipps-University, D-35032 Marburg, Germany}

\date{\today}

\begin{abstract}
  There are several ways to derive Einstein's celebrated formula
  for the energy of a massive particle at rest, $E=mc^2$. Noether's theorem
  applied to the relativistic Lagrange function provides an unambiguous and
  straightforward access to energy and momentum conservation laws but
  those tools were not available at the beginning of the twentieth century and
 are not at hand for newcomers even nowadays.
 In a pedestrian approach, we start from relativistic kinematics and
 analyze elastic and inelastic scattering processes
 in different reference frames to derive the relativistic energy-mass relation.
We extend the analysis
to Compton scattering between a massive particle and a photon,
and a massive particle emitting two photons. Using
the Doppler formula, it follows that
$E=\hbar \omega$ for photons at angular frequency $\omega$
where $\hbar$ is the reduced Planck constant.
We relate our work to other derivations of Einstein's formula
in the literature.
\end{abstract}

\maketitle   

\section{Introduction}
\label{Sec:introduction}

The energy-mass relation for a particle of mass $m$ at rest
($c$: speed of light),
\begin{equation}
  E_0 = m c^2
  \label{eq:emcsquarebare}
\end{equation}
is one of the most popular formulas in physics. It is the basis for
our understanding of the energy production by fusion in stars
and by fission in nuclear power plants.
For this reason, it is desirable to make it accessible to beginners,
not only at university level but preferably already at high-school level.
Consequently, physicists seek to provide an elementary derivation
of the famous formula~(\ref{eq:emcsquarebare}), as are the actual titles of
Einstein's paper in 1935~\cite{Einstein1935} and of
Rohrlich's paper in 1990~\cite{Rohrlich1990}.

The notion of ``elementary derivation'' implies at least three points.
\begin{enumerate}
\item all physical concepts are stated clearly;
\item notions extrinsic to mechanics, e.g., those from
  electrodynamics or even quantum mechanics, are avoided;
\item sophisticated mathematics are kept to a minimum.
\end{enumerate}
Naturally, all derivations must be correct.
This is not always guaranteed, see the comment
by Ruby and Reynolds~\cite{Ruby1991comment}
who pointed out that the non-relativistic Doppler formula used by Rohrlich
is insufficient to derive the relativistic relation~(\ref{eq:emcsquarebare}).
Moreover, the derivations should be self-contained and not use short-cuts
that are justified only a-posteriori.

The second postulation appears impossible to meet because Einstein's formula
invokes the speed of light~$c$. In the preface to the celebrated
textbook
{\sl The Classical Theory of Fields\/}
by Landau and Lifshitz~\cite{Landau-volume2}
where relativity and electromagnetism are treated in one volume,
the authors explicitly state that `A complete, logically connected theory of the
electromagnetic field includes the special theory of relativity,
so the latter has been taken as the basis of the presentation.'
Likewise, Einstein's original considerations~\cite{Einstein1905mc2}
in 1905 are not based on classical mechanics alone.
Einstein addresses a process of emitting electromagnetic waves
by a massive object. He concludes from the energy balance that
when the object loses some amount $\Delta E$ of its energy,
it simultaneously loses the amount $\Delta m = \Delta E/c^2$ of its mass,
\begin{equation}
  \Delta E_0 = \Delta m c^2 \; .
  \label{eq:DeltaE0DeltamEinstein}
\end{equation}
In his arguments, Einstein relies on Maxwell's electrodynamics,
more precisely, to the results from the electrodynamic part
of his famous work
`Zur Elektrodynamik bewegter K\"orper'~\cite{Einstein1905main}.
Also, the relativistic formula for kinetic energy is obtained
by Einstein from electrodynamic arguments, namely, from equations of motion
of a charged particle in an electric field~\cite{Einstein1905main}.
This little historical excursion demonstrates
how closely the relativistic theory is tied to the theory of electromagnetism.

Already in 1906, Planck recognized that the relativistic dynamics
fits into the framework of the principle of least action~\cite{Planck1906};
according to Pais~\cite{Pais-book} this was most probably the first work
on Einstein's special relativity not written by Einstein himself.
Then, in 1907, Minkowski gave a talk in which he introduced four-vectors
in spacetime and showed that the kinetic energy of a massive body is related
to the temporal component of body's four-velocity~\cite{Minkowski1910}.
Thus, very soon after the invention of special relativity
there appeared at least three `points of support' that permit to derive
relativistic dynamics (energy, momentum, etc.)
from relativistic kinematics (time dilation, length contraction,
Lorentz transformations, etc.). These `points of support' remain
electrodynamics, the principle of least action,
Noether's theorem~\cite{Noether1918},
and vectors in Minkowski spacetime.

Albeit of fundamental importance, the concepts introduced
by Planck and Minkowski are much more involved
than relativistic kinematics alone, and could be seen as an obstacle
for beginners who just want to understand relativistic dynamics.
One line of argument that avoids electrodynamics,
the principle of least action, Noether's theorem,
and the notion of four-vectors,
is based on the analysis of particle collisions.
Already in 1909, Lewis and Tolman~\cite{Lewis1909}
used a collision argument to prove the relativistic expression
for the momentum. Their proof is generally adopted in various textbooks,
such as {\sl Spacetime Physics\/}
by Taylor and Wheeler~\cite{Taylor-Wheeler-book},
{\sl The Feynman Lectures on Physics\/}~\cite{Feynman-Lectures-1},
and, in a somewhat restricted version,
in the Berkeley Physics Course~\cite{Berkeley-Course-1}.
Interestingly,
even when they rely on particle collisions,
the authors of textbooks prefer to introduce relativistic
formulas for momentum and energy \emph{ad hoc},
and use thought experiments with collisions only as supporting arguments.
For example, in {\sl Spacetime Physics}~\cite{Taylor-Wheeler-book}
the authors postulate that momentum and energy are just parts
of the four-vector of relativistic momentum,
and write down the corresponding expressions for them.
Only at the end of the chapter, as an exercise,
they provide a derivation of the relativistic
momentum through a collision experiment.
This reflects a `Babylonian' approach to
physics~\cite[p.~47]{Feynman-character-of-physical-law}
rather than the `Greek' method in mathematics'~\cite{Babel}.

In this work we shall employ relativistic kinematics and analyze
two-particle scattering processes to derive the energy-momentum relation
that leads to Einstein's formula~(\ref{eq:emcsquarebare}).
Since we shall analyze scattering processes with very simple geometries,
we require first and second-order Taylor expansion
and the solution of simple first-order differential equations
as mathematical tools to turn the Babylonian approach to an Euclidean one.

Our paper is organized as follows. In Sec.~\ref{Sec:kinematics}
we review the properties of point particles in classical mechanics and
collect the Lorentz transformation and relativistic Doppler formulas.
In Sec.~\ref{Sec:Lagrange}, to set a point of reference,
we briefly derive the dynamics of a single
particle using the principle of least action, and identify
momentum and energy using Noether's
theorem~\cite{Noether1918,Landau-volume2}.
In Sec.~\ref{Sec:pedestrian} we provide the pedestrian derivation
of those formulas using only basic concepts of classical mechanics
as outlined in Sec.~\ref{Sec:kinematics} applied to two-particle scattering.
Using the relativistic Doppler effect, we show in Sec.~\ref{sec:photondyn}
from particle-photon (Compton) scattering
that a photon with angular frequency $\omega$ has the energy $E= \hbar \omega$,
where $\hbar$ is the reduced Planck constant~\cite{Planck1901}.
The same result can be obtained from particle-antiparticle annihilation.
Furthermore, in Sec.~\ref{sec:whatdidtheothersdo}, we briefly
review other approaches to derive Einstein's formula.
Short conclusions, Sec.~\ref{Sec:conclusions}, close our presentations.
Mathematical derivations are deferred to three appendices.

\section{Kinematics: point particle and Lorentz transformation}
\label{Sec:kinematics}

Momentum and energy are concepts of particle dynamics.
Before we address the equations of motions of a single particle
in Sect.~\ref{Sec:Lagrange} and derive Einstein's
formula in Sect.~\ref{Sec:pedestrian},
we first recall the basic concept of a point particle in
classical mechanics. Next, we collect the Lorentz transformation
formulas for the transformation of coordinates
and velocities between two inertial systems.
Since we address photon dynamics in Sect.~\ref{sec:photondyn},
we also collect the formulas for the relativistic Doppler
effect in the present section.

\subsection{Point particle in classical mechanics}

The first axiom of mechanics defines the setting of space-time:
space-time is four-dimensional, i.e., an event is given by a point
$P$ with four coordinates $(t,x,y,z)$ in some (inertial) reference frame.

The second axiom in classical mechanics states
that a particle is at some spatial point $\vecr_1$ at time $t_1$
and arrives at some other spatial point $\vecr_2$ at time $t_2>t_1$ whereby
the world-line that contains all intermediate points
$P(t)=(t_1\leq t\leq t_2,\vecr(t))$
is continuous and (at least) twice differentiable with respect to
the time~$t$.

Kinematics describes the functional dependence of $\vecr(t)$
on the time~$t$.

\subsection{Lorentz transformation}

Apparently, kinematics requires the use of coordinate systems.
However, the choice of a reference system seems to
prefer one coordinate system over the other.
The third axiom of classical mechanics, the Galilean principle of special relativity,
states that this must not be the case:
the equations of motion must have the same functional form
in all inertial reference frames.
A frame that moves with constant velocity~$\vecu$ with respect to
an inertial frame also is an inertial frame. The necessity of inertial frames is
overcome in the theory of general relativity.

What remains unspecified in the axiom is the transformation of coordinates
between two such inertial frames.
Maxwell's equations describe the propagation of light. They
are form-invariant under coordinate transformations
if time and space transform according to the Lorentz transformation formulas.
Lorentz transformations guarantee that
the relativistic distance between two events
is the same in all coordinate systems.
The coordinates of an event $P$ is described
in $K$ with the coordinates $(t,x,y,z)$
and in $K'$ with coordinates $(t',x',y',z')$.
The invariant distance between two events $P_1$ and $P_2$ is given by
\begin{equation}
  s_{12}^2=c^2(t_2-t_1)^2-(x_2-x_1)^2-(y_2-y_1)^2-(z_2-z_1)^2 \;,
\end{equation}
and $s_{12}^2=(s_{12}')^2$ must hold with
\begin{equation}
  (s_{12}')^2=c^2(t_2'-t_1')^2-(x_2'-x_1')^2-(y_2'-y_1')^2-(z_2'-z_1')^2 \; ,
\end{equation}
where we use the coordinates of the two events $P_1$ and $P_2$
in the two different reference frames.
When the two events are infinitesimally close,
$({\rm d}s)^2=c^2({\rm d}t)^2-({\rm d}x)^2-({\rm d}y)^2-({\rm d}z)^2$
is the invariant distance. Note that the velocity of light is the same in all reference
frames, $c'=c$ (Einstein's axiom of the invariance of the speed of light).

To simplify the discussion,
we assume that the two reference frames~$K$ and $K'$
coincide at time $t=t'=0$, and $K'$ moves with velocity~$\vecu=u\vece_x$.
Then, the coordinates of an event $P$ is described
in $K$ with the coordinates $(t,x,y,z)$
and in $K'$ with coordinates $(t',x',y',z')$.
The Lorentz transformation provides the relation between the coordinates,
\begin{equation}
  x =\frac{x'+ut'}{\sqrt{1-u^2/c^2}} \; ,
  \quad y=y' \;, \quad
    z=z' \;, \quad
    t =\frac{t'+ux'/c^2}{\sqrt{1-u^2/c^2}} \; .
    \label{eq:coordinateLorentz}
\end{equation}
For infinitesimal distances, one simply has to replace
$(t,x,y,z)$ by $( {\rm d}t, {\rm d}x, {\rm d}y, {\rm d}z)$
and
$(t',x',y',z')$ by $( {\rm d}t', {\rm d}x', {\rm d}y', {\rm d}z')$.

The velocities of a particle are given by
$\vecv={\rm d}\vecr/({\rm d} t)$ in $K$
  and $\vecv'={\rm d}\vecr'/({\rm d} t')$ in $K'$.
    Therefore,
    they transform according to
\begin{multline}
v_x = \frac{v_x'+u}{1+v_x'u/c^2}\;, \quad
v_y = v_y' \frac{\sqrt{1-u^2/c^2}}{1+v_x'u/c^2}\;,  \\
v_z = v_z' \frac{\sqrt{1-u^2/c^2}}{1+v_x'u/c^2}\; ,
\label{eq:velotransformation}
\end{multline}
when we use the Lorentz transformation
for the coordinates~(\ref{eq:coordinateLorentz})
in their infinitesimal form.

\subsection{Doppler effect}
\label{subsec:Doppler}

Light is described by electromagnetic plane waves
with frequency $\omega$ and wave vector $\veck$
as solutions of the Maxwell equations in the absence of external sources, e.g.,
\begin{equation}
\vecE(\vecr,t) = \vecE_0 \cos(\omega t-\veck\cdot \vecr)
\end{equation}
for the vector of the electric field where $\vecE_0$ is a
three-dimensional vector with real components.
Apparently, $\vecE(\vecr,t)$ has extrema and zeros when the phase
\begin{equation}
  \varphi(\vecr,t) = \omega t-\veck\cdot \vecr
\end{equation}
is a multiple of $\pi/2$. The number of zeros
or the number of maxima/minima are independent of the reference frame
so that the phase must be a Lorentz scalar. This implies that
$(\omega,\veck)$ form a relativistic four-vector in the same way as $(t,\vecr)$
so that its components transform analogously
to eq.~(\ref{eq:coordinateLorentz}),
\begin{multline}
  k_x =\frac{k_x'+u\omega'/c^2}{\sqrt{1-u^2/c^2}} \; ,
  \quad k_y=k_y' \;, \quad
    k_z=k_z' \;, \\
    \omega =\frac{\omega'+uk_x'}{\sqrt{1-u^2/c^2}}
    \label{eq:coordinateLorentzkomega}
\end{multline}
when $K$ and $K'$ move with constant velocity~$\vecu=u\vece_x$ relative
to each other. When the light also travels along the $x$-axis
to the right, we have
$k_y=k_z=0$ and $k_x=k=\omega/c$ from the dispersion relation
\begin{equation}
  \omega =|\veck| c \; .
  \label{eq:dispersionrelation}
\end{equation}
Therefore, we obtain the relativistic Doppler formula for the frequency shift,
\begin{equation}
\omega= \omega' \sqrt{\frac{1+u/c}{1-u/c}}
  \label{eq:Dopplershift}
\end{equation}
between the frequencies measured in $K$ and $K'$.
When the light travels to the left ($k_y=k_z=0$ and $k_x=-k=-\omega/c$),
the signs `$+$' and `$-$' in eq.~(\ref{eq:Dopplershift}) swap their places.

\section{Dynamics: Lagrange formalism}
\label{Sec:Lagrange}

The ultimate goal in classical mechanics is to derive
the motion of particles from basic principles, i.e.,
to formulate equations from which the particle trajectory $\vecr(t)$ can be
deduced. Newton's original formulation was superseded
by the Lagrange and Hamilton formulation
because the underlying Hamilton principle of least action
constitutes the basis of present-day theoretical physics.

\subsection{Particle mass}
\label{subsec:mass}

To describe particle dynamics,
Newton assigns a second defining property to a point particle,
namely its (inertial) mass~$m$.
Below we shall assume that
\begin{enumerate}
\item the non-relativistic momentum
  and (kinetic) energy of a particle are proportional to the mass~$m$;
\item the mass is a scalar under Lorentz transformations;
\item a particle with mass~$M$ can decay into two particles
  with mass $m\leq M/2$.
\end{enumerate}
Property~1 seems self-evident because it
requires twice the force to push two mugs of beer over a counter
compared to pushing a single one. Moreover, we tacitly assume that we
do not gain or lose liquid
when looking at the mug from different reference frames (property~2),
and we know that we can split a liquid into equal volumes
without loosing any (property~3).

On a more fundamental level, the generation of inertial mass requires
an understanding
of the interaction of relativistic particle fields with the Higgs field.
Even more intricate is the notion of a gravitational mass and its equivalence to
the inertial mass. We will not dwell into these fundamental issue here
but move on to the equations of motions that govern the particle dynamics.

\subsection{Euler-Lagrange equations}
\label{subsec:ELequations}

Modern physics is based on the principle of least action.
For a point particle in classic mechanics,
the action~$S$ along a path $\vecR(t)$ with velocity~$\dot{\vecR}(t)$
within the time interval $[t_1,t_2]$ reads
\begin{equation}
  S= \int_{t_1}^{t_2} {\rm d}t L(\vecR,\dot{\vecR},t) \; .
  \label{eq:action}
\end{equation}
Here, $L$ is the Lagrange function that depends only on the particle
coordinates~$\vecR(t)$, velocities~$\dot{\vecR}(t)$, and time~$t$.
Consequently, the particle acceleration must be a function of
the particle position and velocity only,
i.e., the particle motion is deterministic.

To find the realized trajectory $\vecr(t)$, the principle of least action
states that
$S$ is stationary with respect to small variations of the realized trajectory
whereby all trajectories start and end
at the points $P_1=\vecr(t_1)$ and $P_2=\vecr(t_2)$,
respectively. For this reason, the Lagrange function is not unique.
For example, the variation does not change when we add a constant~$C$
to $L$, i.e., using $\tilde{L}=L+C$ in $S$
leads to the same realized trajectory as using $L$.

As shown in textbooks, the realized trajectory $\vecr(t)$
fulfills the Euler-Lagrange equations
\begin{equation}
  \frac{{\rm d}}{{\rm d}t}
  \left. \frac{\partial L}{\partial \dot{\vecR}}
  \right|_{\vecR=\vecr,\dot{\vecR}=\dot{\vecr}}
  = \left. \frac{\partial L}{\partial \vecR}
  \right|_{\vecR=\vecr,\dot{\vecR}=\dot{\vecr}} \; .
    \label{eq:EulerlagrangePP}
\end{equation}
The equations~(\ref{eq:EulerlagrangePP}) constitute Newton's second law.

For example, in non-relativistic mechanics, a single point-particle
has the Lagrange function
\begin{equation}
  L^{\rm nr}(\vecR,\dot{\vecR},t)=\frac{m}{2} \dot{\vecR}^2 \; .
  \label{eq:Lnonrel}
\end{equation}
When inserted in eq.~(\ref{eq:EulerlagrangePP}), the
Euler-Lagrange equations read
\begin{equation}
  \frac{{\rm d}}{{\rm d}t} \left( m \dot{\vecr}\right) = \veczero \; ,
  \label{eq:momconservationpointparticleNR}
\end{equation}
i.e., a free particle moves along a straight line (Newton's first law).

\subsection{Energy and momentum}
\label{subsec:EandpfromL}

One big advantage of the Lagrange formalism over Newton's formulation
of classical mechanics lies in the fact that conserved quantities like
{\em momentum\/} and {\em energy\/} are well defined.
As shown by Noether~\cite{Noether1918}, when the Lagrange function
is invariant under translations in space (time),
there is a conserved quantity called {\em momentum\/} ({\em energy}).
Thus, these objects and their conservation laws result from
the homogeneity of space and time.

The simplest example is the non-relativistic
point particle in Sec.~\ref{subsec:ELequations}
where eq.~(\ref{eq:momconservationpointparticleNR}) expresses
that the momentum
\begin{equation}
  \vecp^{\rm nr}  =m \vecv \; ,
  \label{eq:nonrelmomentum}
\end{equation}
with $\dot{\vecr}=\vecv$ on the realized trajectory,
is a conserved quantity for a non-relativistic free particle, i.e.,
$\vecp^{\rm nr}$ does not change in time.

Apparently, momentum conservation
is based on the fact that the Lagrange function
does not depend on $\vecr$. The theory reflects the fact that
space is homogeneous so that $L$ does not change under translations.
Therefore, for a general Lagrange function $L=L(\dot{\vecR})$
the conserved momentum is defined by
\begin{equation}
  \vecp = \left. \frac{\partial L(\dot{\vecR})}{\partial \dot{\vecR}}
  \right|_{\dot{\vecR}=\dot{\vecr}} \; .
  \label{eq:howtodeterminep}
\end{equation}
We shall derive the relativistic Lagrange function
in the next subsection~\ref{subsec:Action}
and thus readily find the expression for the relativistic momentum
of a point particle.

Homogeneity in time implies that $L$ does not explicitly depend on time,
$L=L(\vecR,\dot{\vecR})$.
The resulting conserved quantity is the energy (or Jacobi integral),
\begin{equation}
  E= \dot{\vecr} \cdot \left.
  \frac{\partial L(\vecR,\dot{\vecR})}{\partial \dot{\vecR}}
  \right|_{\vecR=\vecr,\dot{\vecR}=\dot{\vecr}}
  -L(\vecr,\dot{\vecr})
  \; .
  \label{eq:howtodetermineenergy}
\end{equation}
For the non-relativistic point particle in Sec.~\ref{subsec:ELequations}
we thus find
($\dot{\vecr}=\vecv$)
\begin{equation}
  E^{\rm nr}= \dot{\vecr} \cdot (m \dot{\vecr}) - \frac{m}{2} \dot{\vecr}^2
  = \frac{m}{2} \vecv^2 \equiv T^{\rm nr}   \; ,
  \label{eq:TandEpointparticleNR}
\end{equation}
the well-known expression for the
(kinetic) energy of a non-relativistic point particle.

\subsection{Action for a particle in Minkowski space}
\label{subsec:Action}

Axiomatically, the action~$S$ is a scalar under Lorentz transformations.
The only infinitesimal scalar
for a single particle
on the world line from $P_1$ to $P_2$ is the infinitesimal
distance ${\rm d}s$ between two point on the world-line. Therefore,
\begin{equation}
  S= - \alpha \int_{P_1}^{P_2} {\rm d}s = - \alpha c \int_{t_1}^{t_2} {\rm d}t
  \sqrt{1-\dot{\vecR}^2/c^2} \; .
  \label{eq:relativisticaction}
\end{equation}
Thus, we know the relativistic Lagrange function up to
a constant $\alpha>0$ that is determined from the comparison with the
non-relativistic limit.
Indeed, we can read off the relativistic Lagrange
function from eq.~(\ref{eq:relativisticaction}),
\begin{equation}
  L(\vecR,\dot{\vecR},t)\equiv L(\dot{\vecR})= -\alpha c
  \sqrt{1-\dot{\vecR}^2/c^2} \; .
  \label{eq:Lrelalmost}
\end{equation}
For small velocities $|\dot{\vecR}|\ll c$, the Taylor expansion
leads to
\begin{equation}
L(\dot{\vecR})\approx -\alpha c \left( 1- \frac{\dot{\vecR}^2}{2c^2}
\right)= -\alpha c + \frac{\alpha}{2c} \dot{\vecR}^2 \; .
\end{equation}
The comparison with the non-relativistic Lagrange function
in eq.~(\ref{eq:Lnonrel}) shows that we must set
$\alpha=mc$ to arrive at $L(\dot{\vecR})\approx L^{\rm nr}(\dot{\vecR})+C$.

\subsection{Energy and momentum for a single particle}
\label{subsec:Eandpelegant}

With the relativistic Lagrange function from eq.~(\ref{eq:Lrelalmost}),
\begin{equation}
L(\dot{\vecR})= -m c^2  \sqrt{1-\dot{\vecR}^2/c^2}
  \label{eq:Lrel}
\end{equation}
and the results from Sec.~\ref{subsec:EandpfromL} we can readily
determine the conserved momentum
of a single point particle with mass~$m$, see eq.~(\ref{eq:howtodeterminep}),
\begin{equation}
  \vecp = \frac{m\vecv}{\sqrt{1-\vecv^2/c^2}}
  =\gamma m \vecv
  \label{eq:prelativisticfinal}
\end{equation}
with the relativistic factor
\begin{equation}
\gamma = \frac{1}{\sqrt{1-\vecv^2/c^2}} \; .
  \end{equation}
The particle's conserved energy reads, see eq.~(\ref{eq:howtodetermineenergy}),
\begin{equation}
  E = \vecv \cdot \frac{m\vecv}{\sqrt{1-\vecv^2/c^2}}+ mc^2\sqrt{1-\vecv^2/c^2}
  = \gamma mc^2 \; .
  \label{eq:Energyrelativisticfinal}
\end{equation}
For a particle at rest, $\vecv=\veczero$, the energy is finite,
\begin{equation}
  E_0\equiv E(\vecv=\veczero) = mc^2 \; ,
  \label{eq:Energyrelativisticatrest}
\end{equation}
the famous Einstein formula for the rest energy of a particle.

Equations~(\ref{eq:prelativisticfinal}) and~(\ref{eq:Energyrelativisticfinal})
constitute the main results that need to be proven using
the pedestrian approach
outlined in the next section.

\section{Pedestrian derivation}
\label{Sec:pedestrian}

The derivation in Sect.~\ref{Sec:Lagrange}
is concise and elegant but it uses a number of concepts
in theoretical physics that were not common knowledge at
the beginning of the 20th century nor are they familiar
to newcomers nowadays. For this reason, we collect the main ideas
to derive comprehensively the Einstein formula using only
basic concepts of classical mechanics as outlined in
Secs.~\ref{Sec:kinematics} and~\ref{Sec:Lagrange}.

We pursue the following strategy.
\begin{enumerate}[label={\bf \Alph*.}]
\item Derive the relativistic momentum from elastic scattering of
  two particles;
\item Derive the relativistic kinetic energy from elastic scattering of
two particles;
\item Derive the mass defect formula from fission of a heavy particle
  into two equal light particles;
\item Derive the Einstein energy-mass relation.
\end{enumerate}
We only invoke the concepts of relativistic classical mechanics and do not
refer to classical electrodynamics (electromagnetic waves) or
concepts of quantum mechanics (photons).

\begin{figure}[ht]
  \includegraphics[width=5cm]{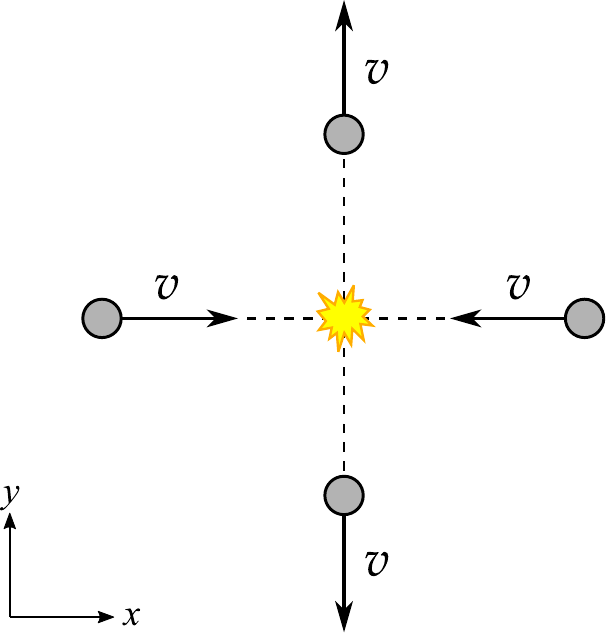}
\caption{Collision of two identical particles with velocities~$\pm \vecv$
in the center-of-mass frame.\label{Fig:particlecollCOM}}
\end{figure}

\subsection{Elastic scattering: momentum conservation}
\label{subsec:Pconservation}

We start our investigation with an elastic scattering of
two identical classical particles, as shown in Fig.~\ref{Fig:particlecollCOM}.
The particles approach each other on the $x$-axis and move
along the $y$-axis after the scattering.
Elastic scattering means that there is no loss of energy
and the particles remain the same so that the speed of each particle
after the impact will remain the same ($v=|\vecv|$),
only its direction will change.
It is intuitively clear that the laws of momentum and energy conservation
are fulfilled: the total momentum is zero before and after the collision,
and the total energy of the two particles remains the same.

We now demand that momentum conservation is also fulfilled in
a different frame of reference.
Let there be two reference frames, one `primed' where
all values will
be marked with primes, the other `unprimed'.
Let the primed frame move relative to the unprimed one in the
direction of the $x$-axis with velocity $\vecu=u \vece_x$.
Then, the connection between the `primed' particle velocity $\vecv'
= (v_x', v_y', v_z')$
and its `unprimed' velocity $\vecv = (v_x, v_y, v_z)$
is given by eq.~(\ref{eq:velotransformation}) in
Sect.~\ref{Sec:kinematics}.

As `primed frame' we choose the center-of-mass frame
as in Fig.~\ref{Fig:particlecollCOM} so that
the `unprimed frame' is the laboratory frame in which the observer
may be viewed at rest. A short calculation convinces the reader that
the momentum of a particle cannot be given by $\vecp^{\rm nr}=m \vecv$
because $m\vecv_1+m\vecv_2\neq m\vecv_3+m\vecv_4$.
Since the system is rotational invariant,
the modulus of the momentum of a particle must be a function
of the modulus of the velocity,
\begin{equation}
|\vecp(\vecv)| = p(v) \; ,
\end{equation}
where $p=\sqrt{p_x^2+p_y^2+p_z^2}$ and $v=\sqrt{v_x^2+v_y^2+v_z^2}$
so that
\begin{equation}
  p_x= \frac{v_x}{v}  p(v)\;,  \quad
p_y = \frac{v_y}{v} p(v) \; , \quad
p_z =\frac{v_z}{v}p(v) \; ,
\end{equation}
with the Cartesian components, $v_{x,y,z}=\vece_{x,y,z}\cdot \vecv$,
 $p_{x,y,z}=\vece_{x,y,z}\cdot \vecp$.

\begin{figure}[ht]
  \includegraphics[width=\linewidth]{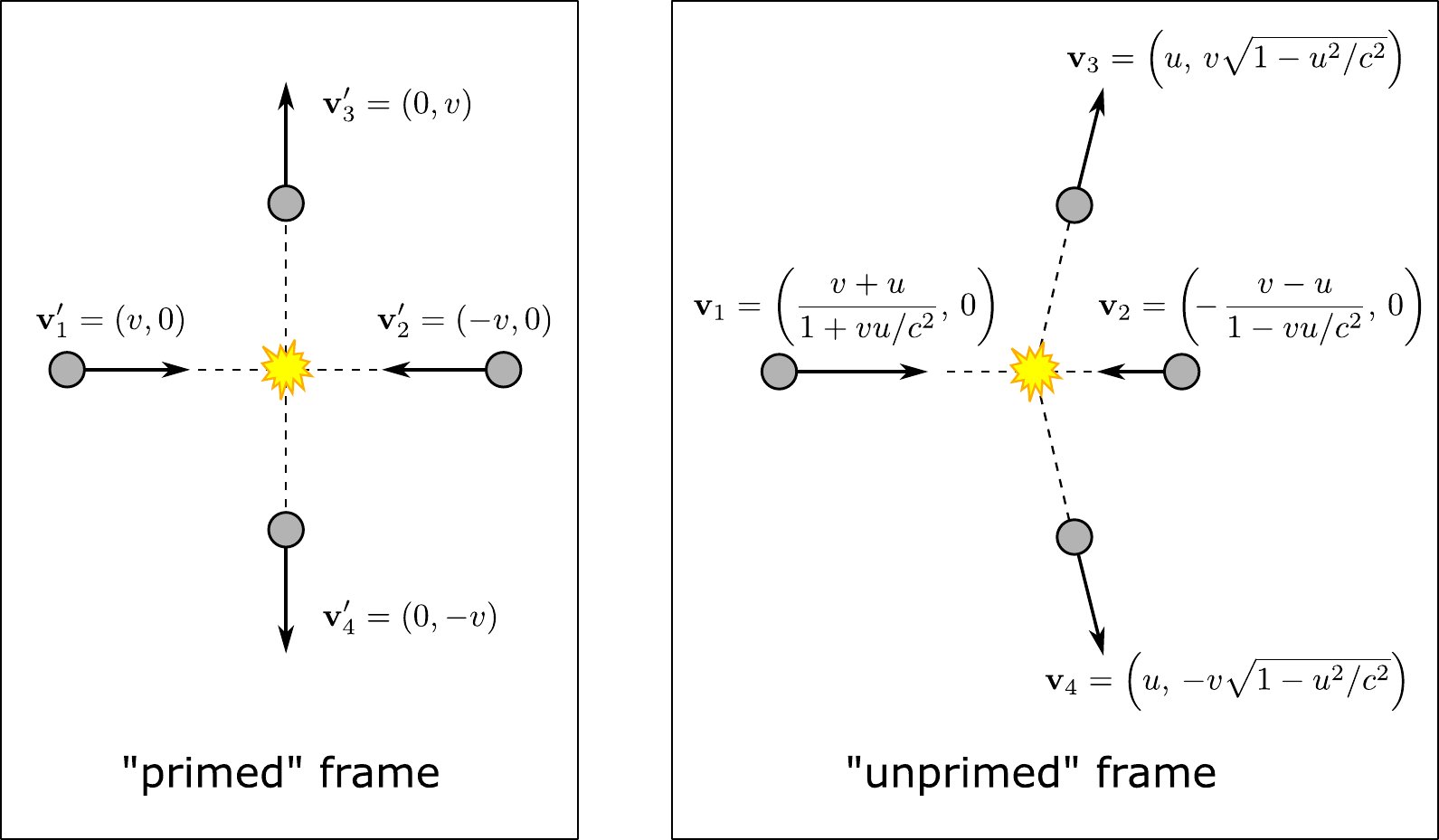}
\caption{Elastic collision in the `primed' center-of-mass
  reference frame (left)
  and in the `unprimed' laboratory reference frame (right).
  The `primed' frame moves to the right with velocity $\vecu=u\vece_x$,
  relative to the `unprimed' one.\label{Fig:bothframes}}
\end{figure}

To find the unknown function $p(v)$, we use the conservation of
momentum in a collision,
\begin{equation}
\vecp_1 + \vecp_2 = \vecp_3 + \vecp_4
\end{equation}
for which we only need to consider the $x$-projection of this equality
because the other components just give $0=0$.
We consider the elastic scattering event in
the laboratory frame, see Fig.~\ref{Fig:bothframes}.
We note that
\begin{equation}
p_{1,x} = p_1 = p(v_1) = p\left( \frac{v+u}{1+vu/c^2} \right)
\end{equation}
because the velocity $\vecv_1$ is just directed along the $x$-axis.
Next,
\begin{equation}
p_{2,x} = -p_2 = -p(v_2) = -p\left( \frac{v-u}{1-vu/c^2} \right) \; .
\end{equation}
To be definite, we assume that $u$ is less than $v$
so that the velocity $\vecv_2$ is directed towards the negative $x$-axis,
and hence $p_{2,x} = -p_2$.
Lastly,
\begin{multline}
  p_{3,x} = \frac{v_{3,x}}{v_3} p(v_3) =\\
  = \frac{u}{\sqrt{v^2+u^2-v^2u^2/c^2}}
  p\left( \sqrt{v^2+u^2-v^2u^2/c^2} \right) \; ,
\end{multline}
and $p_{4,x} = p_{3,x}$.
When we insert these expressions into
$p_{1,x} + p_{2,x} = p_{3,x} + p_{4,x}$, we obtain
\begin{multline}
  p\left( \frac{v+u}{1+vu/c^2} \right)
  - p\left( \frac{v-u}{1-vu/c^2} \right) = \\
  = 2\frac{u}{\sqrt{v^2+u^2-v^2u^2/c^2}}
  p\left( \sqrt{v^2+u^2-v^2u^2/c^2} \right) \; .
  \label{eq:definingeqforpofv}
\end{multline}
This is the equation for the unknown function $p(v)$.
The unique solution for $p(v)$ for {\em all\/}~$0<u<v<c$ is
\begin{equation}
  p(v) = C_p \frac{v}{\sqrt{1-v^2/c^2}}
  \label{eq:pofvwithconstant}
\end{equation}
with some constant $C_p$ which
is readily achieved after some algebra, or by using
{\sc Mathematica}~\cite{Mathematica12}.
A derivation of eq.~(\ref{eq:pofvwithconstant}) is given
in appendix~\ref{app:givepofv}.

It remains to determine the constant~$C_p$
in eq.~(\ref{eq:pofvwithconstant}).
It is known that, for small velocities, $|\vecv| \ll c$,
we have the non-relativistic
relation $\vecp^{\rm nr}=m\vecv$ for the relation between
particle velocity and its momentum, see eq.~(\ref{eq:nonrelmomentum}).
The expansion of eq.~(\ref{eq:pofvwithconstant}) for small $v$ gives
$p(v\ll c)\approx C_p v$ so that the constant
is the mass of the particle, $C_p=m$.
Thus, $p(v) = mv/\sqrt{1-v^2/c^2}$, $p = \gamma mv$, or, in vector form,
\begin{equation}
  \vecp = \gamma m \vecv \; .
  \label{eq:relpagain}
\end{equation}
In sum, we re-derived the relativistic momentum
as a function of the particle velocity as
given in eq.~(\ref{eq:prelativisticfinal}).

\subsection{Elastic scattering: kinetic energy conservation}
\label{subsec:Ekinconservation}

We reconsider the scattering experiment in Figs.~\ref{Fig:particlecollCOM}
and~\ref{Fig:bothframes}
in the context of energy conservation.
The collision in this experiment is {\em elastic}, i.e.,
the internal states of the particles do not change.
The only difference between the states of the particles
results from their velocities. Therefore,
the energy $E$ of each particle is a function of its velocity $v$ only,
$E = E(v)$, again assuming rotational invariance.
Energy conservation thus implies
\begin{equation}
E(v_1) + E(v_2) = E(v_3) + E(v_4)
\end{equation}
in the scattering event in both reference frames.

It is convenient to divide the particle's energy $E$ into
its internal energy (also called `potential energy' or `rest energy') $E_0$,
and its kinetic energy $T$,
\begin{equation}
  E(v) = E_0 + T(v)\; .
  \label{eq:totalE}
\end{equation}
The internal energy is nothing else but the energy at zero velocity,
$E_0 \equiv E(0)$.
Since the internal energy does not change in elastic collisions,
we can rewrite the energy conservation law for our scattering
experiment as conservation of the kinetic energy,
\begin{equation}
  T(v_1) + T(v_2) = T(v_3) + T(v_4)\; .
  \label{eq:conservedkineticenergy}
\end{equation}
Apparently, the value of $E_0$ cannot be determined using
elastic scattering.

Now, we derive the function $T(v)$ from
the condition~(\ref{eq:conservedkineticenergy}).
In the `primed' reference frame (center-of-mass system, left part of
Fig.~\ref{Fig:bothframes}),
it is evident that the kinetic energy is conserved,
$T(v_1') + T(v_2') = T(v_3') + T(v_4')$, because
the velocity moduli are all equal,
$v_1'=v_2'=v_3'=v_4'=v$.

For the laboratory frame, things are a little bit more complicated.
The velocity moduli are
\begin{multline}
v_1 = \frac{v+u}{1+vu/c^2}\;, \quad
v_2 = \frac{v-u}{1-vu/c^2}\;, \\
v_3 = v_4 = \sqrt{v^2 + u^2 - \frac{v^2u^2}{c^2}}\; .
\label{eq:unprimedvelocities}
\end{multline}
As for the case of the momentum, the
non-relativistic expression, $T^{\rm nr}(v)= mv^2/2$
in eq.~(\ref{eq:TandEpointparticleNR}), does not fulfill
eq.~(\ref{eq:conservedkineticenergy}) when
eq.~(\ref{eq:unprimedvelocities}) is used.

We find the correct expression for $T(v)$ as for the momentum function $p(v)$.
For general $0<u<v<c$, we must solve
eq.~(\ref{eq:conservedkineticenergy}) for the velocities given
in eq.~(\ref{eq:unprimedvelocities}),
\begin{multline}
T\left(\frac{v+u}{1+vu/c^2}\right) + T\left(\frac{v-u}{1-vu/c^2}\right)=\\
= 2T\left(\sqrt{v^2 + u^2 - \frac{v^2u^2}{c^2}}\right) \; .
\label{eq:solvethiseqforT}
\end{multline}
The unique solution that reproduces the non-relativistic limit,
$T(v\ll c)\approx mv^2/2$, obeys
\begin{equation}
  T(v)= (\gamma -1)mc^2 \; .
  \label{eq:solutionTofv}
\end{equation}
This is shown explicitly in appendix~\ref{app:prooftofv}.
The proof that $T(v)$ from eq.~(\ref{eq:solutionTofv})
requires only some basic algebra, or can be done using
{\sc Mathematica}~\cite{Mathematica12}.

Eq.~(\ref{eq:solutionTofv}) provides
the relativistic formula for the kinetic energy $T(v)$
of a particle with mass $m$ and velocity~$|\vecv|=v$.
We recall that the mass is considered as independent of the velocity.
The velocity $v$ appears here in the $\gamma$-factor only,
$\gamma = 1/\sqrt{1-v^2/c^2}$.

\subsection{Particle fission: mass defect}
\label{subsec:massdefect}

\begin{figure}[ht]
\includegraphics[width=\linewidth]{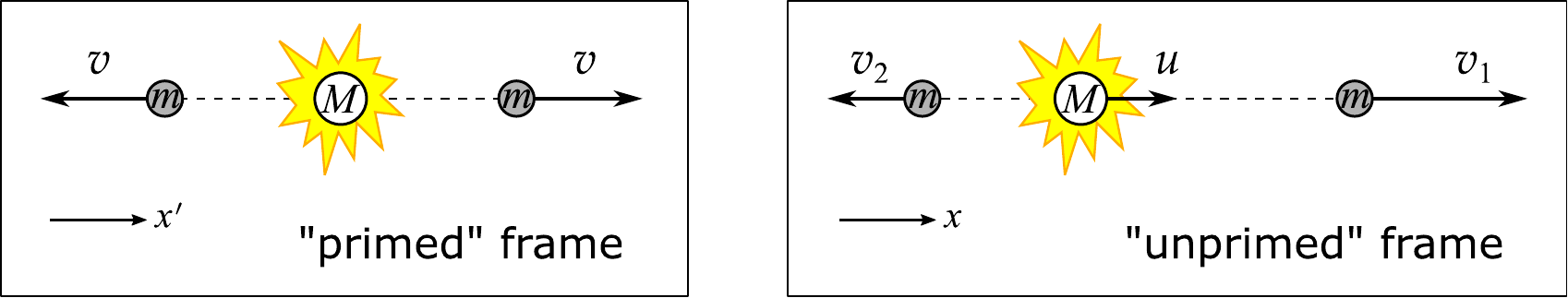}
\caption{Decay of a particle of mass $M$
  into two identical particles of mass $m\leq M/2$.
  The left part depicts this process in the (primed)
  center-of-mass reference frame,
  and the right part in the unprimed frame,
  in which the center of mass moves to the right with velocity~$u$
  (laboratory frame).\label{fig:fission}}
\end{figure}

Thus far, we used elastic collisions as a convenient tool
to derive the relativistic formulas for the momentum and the kinetic energy.
But elastic processes, by definition, do not change internal states
of particles. Therefore, we need to address {\em inelastic\/} processes
to get access to the internal energy.
The simplest example process is the fission of a particle of mass~$M$
into two lighter identical particles of mass~$m\leq M/2$.
This process is shown in Fig.~\ref{fig:fission}
in two different reference frames. When we reverse the arrow of time,
the process describes particle fusion.

The left part of Fig.~\ref{fig:fission}
shows the particle decay in the center-of-mass frame.
The particle with mass $M$ splits into two particles that
fly away in opposite directions with velocities~$\vecv= \pm v \vece_x$.
Before the decay, the kinetic energy of the heavy particle is equal to zero.
After the decay, the total kinetic energy is equal to $2mc^2(\gamma-1)$,
where $\gamma = 1/\sqrt{1-v^2/c^2}$.
Hence, due to energy conservation,
the total internal energy decreases by the same amount,
\begin{equation}
  \Delta E_0(m) =  2mc^2(\gamma-1)\; .
  \label{eq:E0decrease}
\end{equation}
This decrease in internal energy is proportional
to the mass defect as we shall show next.

To this end, we apply the momentum conservation
law in the unprimed (laboratory) frame.
The right part of Fig.~\ref{fig:fission}
shows the same fission process in a reference frame, where
the center-of-mass frame moves to the right with velocity $u<v<c$.
In this frame, the heavy particle moves to the right with velocity~$u$,
one light particles moves to the right with velocity
\begin{equation}
  v_1 = \frac{v+u}{1+vu/c^2}\; ,
  \label{eq:v1fission}
\end{equation}
and the other light particle moves
to the left with velocity
\begin{equation}
  v_2 = \frac{v-u}{1-vu/c^2}\; .
    \label{eq:v2fission}
\end{equation}
The momentum conservation law along the $x$-axis reads
\begin{equation}
  \frac{Mu}{\sqrt{1-u^2/c^2}} = \frac{mv_1}{\sqrt{1-v_1^2/c^2}}
  - \frac{mv_2}{\sqrt{1-v_2^2/c^2}} \; .
\end{equation}
Therefore, the heavy mass reads
\begin{equation}
  M =  m \frac{\sqrt{1-u^2/c^2}}{u} \left( \frac{v_1}{\sqrt{1-v_1^2/c^2}}
  - \frac{v_2}{\sqrt{1-v_2^2/c^2}} \right) \; .
\end{equation}
We substitute $v_{1,2}$ from eqs.~(\ref{eq:v1fission}) and~(\ref{eq:v2fission})
and find after some algebra
\begin{equation}
M =  \frac{2m}{\sqrt{1-v^2/c^2}} \; .
\end{equation}
We see that the mass~$M$ of the heavy particle is
larger than the total mass of the debris particles~$2m$,
assuming that the decay particles move away from each other, $v>0$.
Due to the decay, the total mass decreases by
\begin{equation}
  \Delta M(m) = M - 2m = 2m \left(\frac{1}{\sqrt{1-v^2/c^2}}-1\right)
  =2m(\gamma-1)\; ,
  \label{eq:massdefect}
\end{equation}
which is sometimes called the {\em mass defect}.

Now we can compare the mass defect $\Delta M(m)$ in eq.~(\ref{eq:massdefect})
with the decrease of internal energy $\Delta E_0(m)$ in
eq.~(\ref{eq:E0decrease}), which leads to
\begin{equation}
  \Delta E_0(m) = \Delta M(m) c^2 \; .
  \label{eq:DeltaE0Deltam}
\end{equation}
Apparently, every change of the total internal energy $\Delta E_0(m)$
of the system is accompanied by a change of its total mass $\Delta M(m)$,
such that eq.~(\ref{eq:DeltaE0Deltam}) holds.

\subsection{Energy-mass formula}
\label{subsec:annihilation}

A point particle has no internal degrees of freedom like, e.g., a molecule.
Point particles in classical physics are solely characterized by their mass.
Therefore, its internal energy can only depend on~$m$, $E_0\equiv E_0(m)$.
We apply this Ansatz to eq.~(\ref{eq:E0decrease}) and write
explicitly
\begin{equation}
\Delta E_0(m)=  E_0(M)-2E_0(m) = 2mc^2(\gamma-1) \; .
\end{equation}
Note that the velocity~$v$ was never specified because, in a real
experiment, it is measured and used to interpret
the fission process and the internal structure of the particle of mass~$M$.

In our thought experiment, we can consider the extreme case
that the heavy particle just splits
into two halves that do not separate from each other,
$v=0$. Therefore, the internal energy must obey the relation
\begin{equation}
  \Delta E_0(m)=  E_0(2m)-2E_0(m)=0
  \label{eq:nomassdefectcase}
\end{equation}
because there is no mass defect in this case, $M=2m$, as $\Delta M(m)=0$, see
eq.~(\ref{eq:massdefect}).
Since eq.~(\ref{eq:nomassdefectcase}) must hold for all~$m$, it follows that
\begin{equation}
E_0(m)= C_0 m \; ,
\end{equation}
and the constant must be $C_0=c^2$ in view of eq.~(\ref{eq:DeltaE0Deltam}).
Therefore, we finally arrive at the desired result
\begin{equation}
E_0(m)= m c^2
\end{equation}
for the internal energy of a point particle of mass~$m$.

The total energy for a moving particle is then
\begin{equation}
E(v)= \gamma m c^2 =\frac{mc^2}{\sqrt{1-v^2/c^2}} \; ,
\end{equation}
according to eqs.~(\ref{eq:totalE}) and~(\ref{eq:solutionTofv}),
and in agreement with the expression from the Lagrange formalism
outlined in Sec.~\ref{Sec:Lagrange}, see eq.~(\ref{eq:Energyrelativisticfinal}).

\section{Photon dynamics}
\label{sec:photondyn}

So far we considered collisions of massive particles.
In this section, we study the collision of a massive particle with a photon,
i.e.,
Compton scattering, and the emission of two photons by a massive particle.

\subsection{Compton scattering}
\label{subsec:Comptonscattering}

First, we show that energy and momentum conservation
in the Compton scattering event provides enough information
to prove
\begin{equation}
  E_{\rm ph}=\hbar \omega  \quad, \quad E_{\rm ph}=|\vecp_{\rm ph}| c
  \label{eq:Emcsqforphotons}
\end{equation}
for photons. In this sense, particle scattering alone also proves
Einstein's second famous formula~(\ref{eq:Emcsqforphotons}),
without resorting to
the photoelectric effect.

Our derivation is based on the assumption
that the photon energy is some (yet unknown) function~$f(\omega)$
of its (angular) frequency~$\omega$,
and that the absolute value of its momentum is given by another
function~$g(\omega)$,
\begin{equation}
E_{\rm ph}(\omega)=f(\omega) \quad, \quad |\vecp_{\rm ph}| = g(\omega) \; .
\end{equation}
The latter also depends only on $\omega$ due to the
dispersion relation~(\ref{eq:dispersionrelation}).

For simplicity, we direct the photon momentum along its wave vector.
As in the previous sections, we imply that the functions $f(\omega)$
and $g(\omega)$
permit Taylor expansions at any $\omega>0$.

\begin{figure*}[ht]
\includegraphics[width=\linewidth]{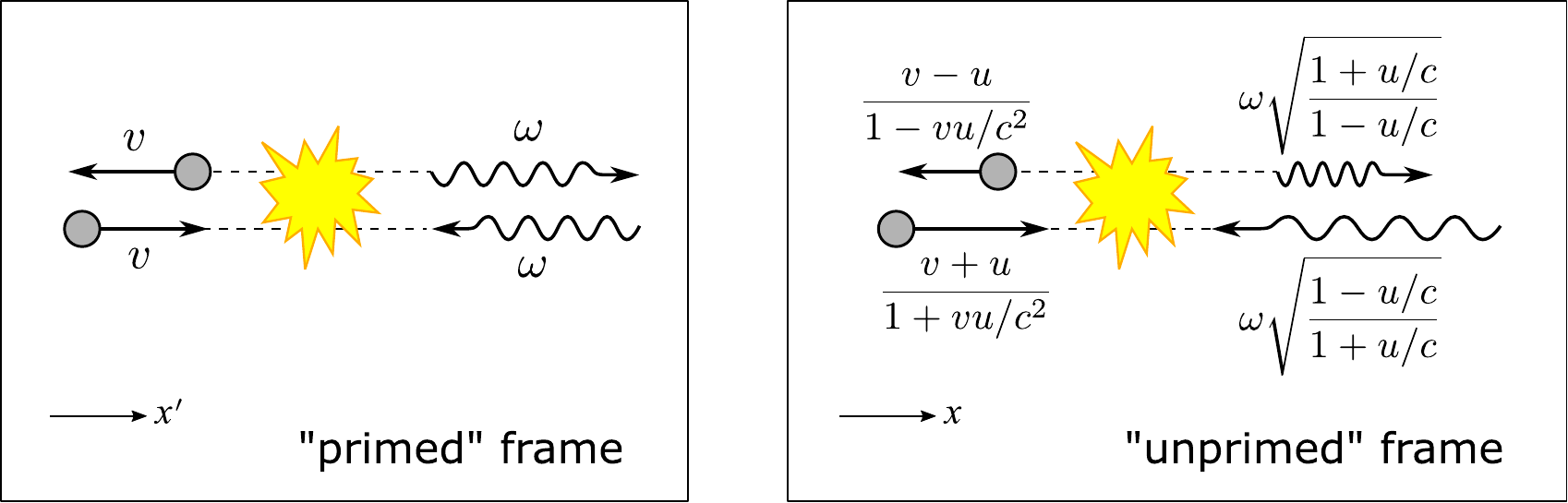}
\caption{Compton scattering process as seen in the center-of-mass frame $K'$
  (left part) and in another reference frame $K$
  that moves to the left relative to $K'$
  with velocity $u$ (right part).\label{fig:compton}}
\end{figure*}

Let us consider a collision of a photon with a massive particle (an electron
for brevity)
in the center-of-mass frame where the net momentum is equal to zero.
To be definite, we let the particles travel towards each other
along the $x$-axis before
the collision, the electron to the right with some velocity $\vecv = v \vece_x$,
and the photon to the left.
The condition of zero net momentum relates the electron velocity $v$
to the photon frequency $\omega$
\begin{equation} \label{eq:compton-v-omega}
\gamma(v) m v = g(\omega) \; ,
\end{equation}
where $m$ is the electron mass, and $\gamma(v) = (1-v^2/c^2)^{-1/2}$
as in Sect.~\ref{Sec:pedestrian}.
The conservation laws permit that
both particle fly apart also along the $x$ axis after the collision,
the electron to the left with velocity $-v \vece_x$,
and the photon to the right with the same frequency $\omega$
as before the collision. This process is schematically depicted
in the left part of Fig.~\ref{fig:compton}.

Now we look at this very process from
another reference frame $K$, see Fig.~\ref{fig:compton}, right part, that
moves to the left along $x$-axes with some velocity $u$,
as seen from the center-of-mass frame $K'$. In the frame $K$,
the initial and the final electron velocities are
\begin{equation}
  \vecv_1=\frac{v+u}{1+vu/c^2}\,\vece_x \quad \text{and} \quad
  \vecv_3= -\frac{v-u}{1-vu/c^2}\,\vece_x \; ,
\end{equation}
correspondingly, as follows from the relativistic velocity addition
rules~(\ref{eq:velotransformation}).
The photon frequency undergoes the Doppler shift when switching from frame $K'$
to frame $K$, see eq.~(\ref{eq:Dopplershift}),
\begin{equation}
  \omega_2= \omega \sqrt{\frac{1-u/c}{1+u/c}} \quad \text{and} \quad
  \omega_4= \omega \sqrt{\frac{1+u/c}{1-u/c}}
\end{equation}
before and after the collision, respectively.

The energy conservation law in the laboratory frame $K$ therefore reads
\begin{multline}
  \gamma\left(\frac{v+u}{1+vu/c^2}\right) mc^2
  + f\left(\omega \sqrt{\frac{1-u/c}{1+u/c}}\right) = \\
  =\gamma\left(\frac{v-u}{1-vu/c^2}\right) mc^2
  + f\left(\omega \sqrt{\frac{1+u/c}{1-u/c}}\right) \; .
  \label{eq:compton-energy-conservation}
\end{multline}
Similarly, the momentum conservation law in the frame $K$
has the following form when projected onto the $x$-axis
\begin{equation*}
  L(v,u,\omega) = R(v,u,\omega) \; , 
\end{equation*}
\begin{multline*}
  L(v,u,\omega) = \\
  = \gamma\left(\frac{v+u}{1+vu/c^2}\right)
  m\,\frac{v+u}{1+vu/c^2} - g\left(\omega \sqrt{\frac{1-u/c}{1+u/c}}\right) \; ,
  \nonumber \\
\end{multline*}
\begin{multline}
R(v,u,\omega) = \\
 = -\gamma\left(\frac{v-u}{1-vu/c^2}\right) m\,\frac{v-u}{1-vu/c^2}
+ g\left(\omega \sqrt{\frac{1+u/c}{1-u/c}}\right) \; .
 \label{eq:compton-momentum-conservation}
\end{multline}
Equations~(\ref{eq:compton-energy-conservation})
and~(\ref{eq:compton-momentum-conservation}) must be fulfilled
for all positive values $\omega$ and for all values of $u$,
provided that electron velocity $v$ is related to $\omega$
by eq.~(\ref{eq:compton-v-omega}).
They permit to determine the functions $f(\omega)$ and $g(\omega)$,
up to a few parameters.

As in the previous section, we expand
eqs.~(\ref{eq:compton-energy-conservation})
and~(\ref{eq:compton-momentum-conservation})
in a power series in~$u$ which provides differential equations
for the functions $f(\omega)$ and $g(\omega)$, see Appendix~\ref{app:compton}.
Their solution leads
to the following expressions for the photon energy $f(\omega)$,
\begin{equation}
  f(\omega) = C_1 \omega - \frac{C_2}{\omega} + C_3 \; ,
  \label{eq:f-omega-via-C}
\end{equation}
and for the photon momentum $g(\omega)$,
\begin{equation}
  g(\omega) = \left( C_1 \omega + \frac{C_2}{\omega} \right) \frac{1}{c} \; ,
  \label{eq:g-omega-via-C}
\end{equation}
where $C_1$, $C_2$, and $C_3$ are some constants.
To fix these constants, we note that the
energy and pressure of light are always positive,
thence $f(\omega)>0$ and $g(\omega)>0$ for all positive frequencies $\omega$.
Therefore,
\begin{equation}
C_1 > 0, \quad C_2 = 0, \quad \text{and} \quad C_3 \ge 0\; .
\end{equation}
Also, knowing that photons can have arbitrarily small energies,
we conclude that $C_3 = 0$. Hence, the only free constant is $C_1$ so that
from $f(\omega)=c g(\omega)$ we find that
\begin{equation}
 E_{\rm ph}=|\vecp_{\rm ph}| c
  \label{eq:Emcsqforphotonshalfway}
\end{equation}
holds. Moreover, the result
\begin{equation}
  E_{\rm ph}(\omega)=C_1 \omega
  \label{eq:almostdoneEhbaromega}
\end{equation}
is nothing but Planck's fundamental assertion that light comes in quanta
(called photons) with energy $E_{\rm ph}(\omega)=\hbar \omega$, i.e.,
the constant $C_1$ was first named by Planck, $C_1=\hbar$.
Using this identification, eqs.~(\ref{eq:Emcsqforphotonshalfway})
and~(\ref{eq:almostdoneEhbaromega})
prove eq.~(\ref{eq:Emcsqforphotons}) entirely.

\subsection{Two-photon emission}
\label{subsec:twophotonemission}

\begin{figure}[ht]
\includegraphics[width=\linewidth]{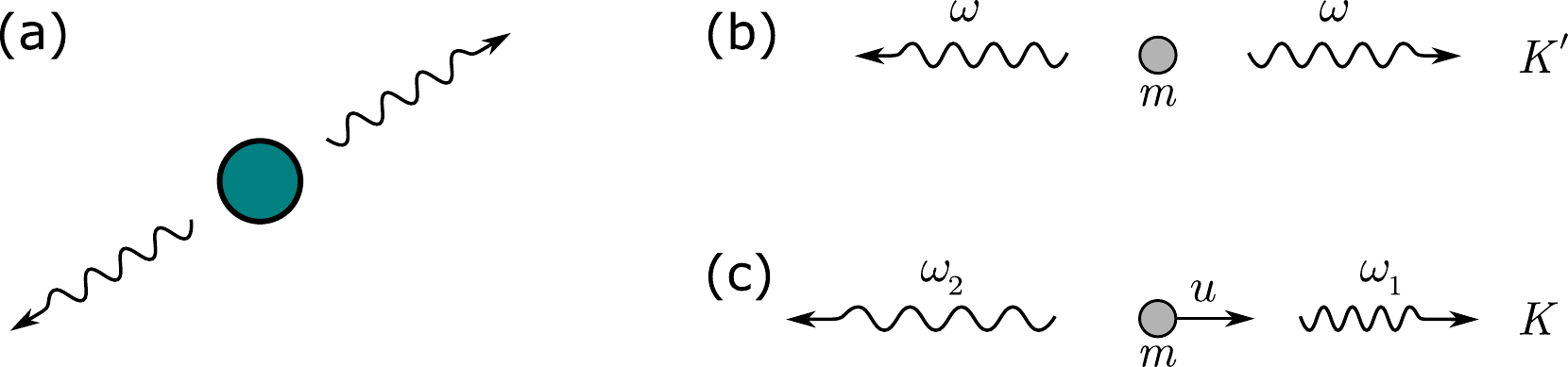}
  \caption{(a) Einstein's seminal thought experiment~\cite{Einstein1905mc2}
    modified later by Feigenbaum and Mermin~\cite{Feigenbaum1988}
    and by Rohrlich~\cite{Rohrlich1990}; (b, c) its simplified version,
    where a particle annihilates into two photons.\label{fig:Einstein}}
\end{figure}

An alternative approach to the mass-defect formula~(\ref{eq:massdefect})
modifies Einstein's original thought experiment~\cite{Einstein1905mc2}
by making it independent of the knowledge of electrodynamics.
In the original setting, a body at rest with mass~$M$
emits two equal bunches of light
in opposite directions, see Fig.~\ref{fig:Einstein}a.
Rohrlich~\cite{Rohrlich1990} argues that even the nineteen-century
non-relativistic physics,
supplied by the photons' energy and momentum formulas,
makes it possible to derive the mass-defect formula~(\ref{eq:massdefect})
by an analysis of the experiment depicted in Fig.~\ref{fig:Einstein}a.
Feigenbaum and Mermin~\cite{Feigenbaum1988} suggest
to replace the light by massive particles
which makes the problem purely mechanical, and we are left with
the problem analyzed in Sec.~\ref{subsec:massdefect}.

Here, we retrace the thought experiment of particle annihilation
into two photons. Note that we make use of the information that
a photon of frequency $\omega$ has energy $E= \hbar\omega$ and
momentum $|\vecp|= \hbar\omega/c$.

In the frame $K'$ where the particle is at rest,
see Fig.~\ref{fig:Einstein}b,
both photons must have the same absolute value of momentum and, consequently,
the same frequency~$\omega$. The energy balance in~$K'$ therefore reads
\begin{equation}
 E_0 = 2\hbar\omega \; ,
  \label{eq:2photons-1}
\end{equation}
where $E_0$ is the particle's internal energy, and $\hbar\omega$
is the energy of one photon.
In another frame $K$, see Fig.~\ref{fig:Einstein}c,
that moves with speed $u$ to the left relative to $K'$,
the photons undergo the Doppler shift and acquire the frequencies
\begin{equation}
 \omega_1 = \omega \sqrt{\frac{1+u/c}{1-u/c}} \quad \text{and} \quad
 \omega_2 = \omega \sqrt{\frac{1-u/c}{1+u/c}} \; ,
  \label{eq:2photons-2}
\end{equation}
according to eq.~(\ref{eq:Dopplershift}).
Next, we apply momentum conservation in the frame $K$.
The particle's momentum $\gamma(u) mu$ turns into the photon momenta,
$\hbar\omega_1/c$ and $\hbar\omega_2/c$, which are directed to the right
and to the left, respectively. Therefore,
\begin{equation}
 \gamma(u) \, mu =
 \frac{\hbar\omega}{c} \sqrt{\frac{1+u/c}{1-u/c}}
 - \frac{\hbar\omega}{c} \sqrt{\frac{1-u/c}{1+u/c}} \; .
  \label{eq:2photons-3}
\end{equation}
We divide both sides by $\gamma(u) = 1/\sqrt{(1+u/c)(1-u/c)}$
and obtain
\begin{equation}
 mu = \frac{2 \hbar \omega u}{c^2} \; .
  \label{eq:2photons-4}
\end{equation}
This gives $\hbar \omega= mc^2/2$ so that substituting into
eq.~(\ref{eq:2photons-1}) finally gives
\begin{equation}
 E_0 = mc^2 \; .
  \label{eq:2photons-5}
\end{equation}
Note that, instead of the momentum conservation in frame $K$,
we can also consider the energy conservation in this frame
which leads to the same conclusion.

\section{Relation to other work}
\label{sec:whatdidtheothersdo}

In the literature, one can find many clever designs that aim at the
construction or derivation of the relativistic momentum,
the kinetic energy, or the mass-to-energy relation from thought experiments
with collisions, including a version by Einstein himself,
dated from the year~1935~\cite{Einstein1935}.
Many of them are cited and discussed by Hu~\cite{Hu2009}
who also suggests two additional collision schemes.
Here, we briefly relate to those that employ particle collisions.

\subsection{Relativistic momentum}
\label{subsec:relmomexamples}

\begin{figure}[b]
\includegraphics[width=\linewidth]{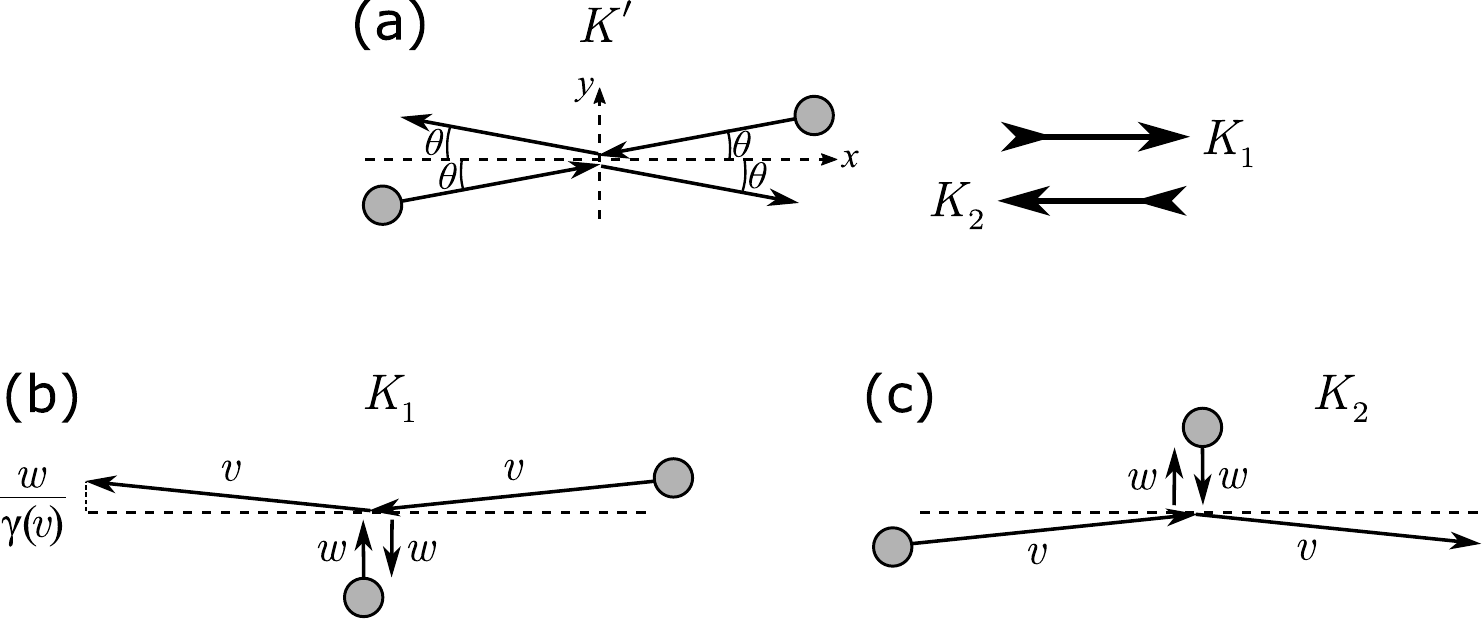}
  \caption{Thought experiment by Lewis and
    Tolman~\cite{Lewis1909,Taylor-Wheeler-book,Feynman-Lectures-1}
    that proves the relativistic formula for the momentum,
    eq.~(\ref{eq:prelativisticfinal}).\label{fig:Lewis}}
\end{figure}

Lewis and Tolman~\cite{Lewis1909} consider
the elastic collision of two identical particles that approach each other
at an infinitely small angle $\theta$ relative to the $x$-axis,
and move at the same angle $\theta$ after the collision,
see Fig.~\ref{fig:Lewis}a.
When we look at this process from the reference frame $K_1$ that
moves to the right relative to the center-of-mass frame $K'$
along with the lower particle, see Fig.~\ref{fig:Lewis}b,
one recognizes that the lower particle moves up and down at some small,
non-relativistic speed~$w$. The higher particle moves at
some speed~$v$. In the frame $K_2$ that
moves to the left along with the upper particle, see Fig.~\ref{fig:Lewis}c,
the particles exchange their roles:
the upper one moves at the low speed~$w$,
and the lower one at the high speed~$v$.
The comparison of Figs.~\ref{fig:Lewis}b and~\ref{fig:Lewis}c shows
that the vertical projection of the upper particle's velocity
in frame $K_1$ is equal to $\pm w/\gamma(v)$,
where the denominator arises from the relativistic time dilation.
Therefore the momentum conservation law in the projection
to the vertical axis reads
\begin{equation}
  mw - p(v)  \frac{w/\gamma(v)}{v} = - mw + p(v) \, \frac{w/\gamma(v)}{v} \, ,
  \label{eq:Lewis1}
\end{equation}
where $m$ is the particle mass.
Resolving this equation with respect to the particle's momentum $p(v)$,
one readily recovers the relativistic formula $p(v) = \gamma(v) mv$,
eq.~(\ref{eq:prelativisticfinal}).

A number of schemes use the following convenient rule for the transformation
of the $\gamma$-factor between the `primed' and `unprimed' reference frames,
\begin{equation}
  \gamma(v) = \gamma(v') \gamma(u)
  \left( 1 + \frac{\vecv'\cdot\vecu}{c^2} \right) \; ,
  \label{eq:gamma-rule}
\end{equation}
where $\vecu$ is the velocity of the `primed' frame relative
to the `unprimed' one. Eq.~(\ref{eq:gamma-rule}) is a direct consequence
of the velocity transformation rule~(\ref{eq:velotransformation}).
Using eq.~(\ref{eq:gamma-rule}) one can rewrite the transformation rules
for the components of the velocity in the $y,z$-directions,
assuming that $\vecu$ is parallel to the $x$-axis,
\begin{equation}
  v_y = v'_y \, \frac{\gamma(v')}{\gamma(v)} \; , \quad
  v_z = v'_z \, \frac{\gamma(v')}{\gamma(v)} \; .
  \label{eq:vy-vz-rule}
\end{equation}

\begin{figure}[ht]
\includegraphics[width=0.7\linewidth]{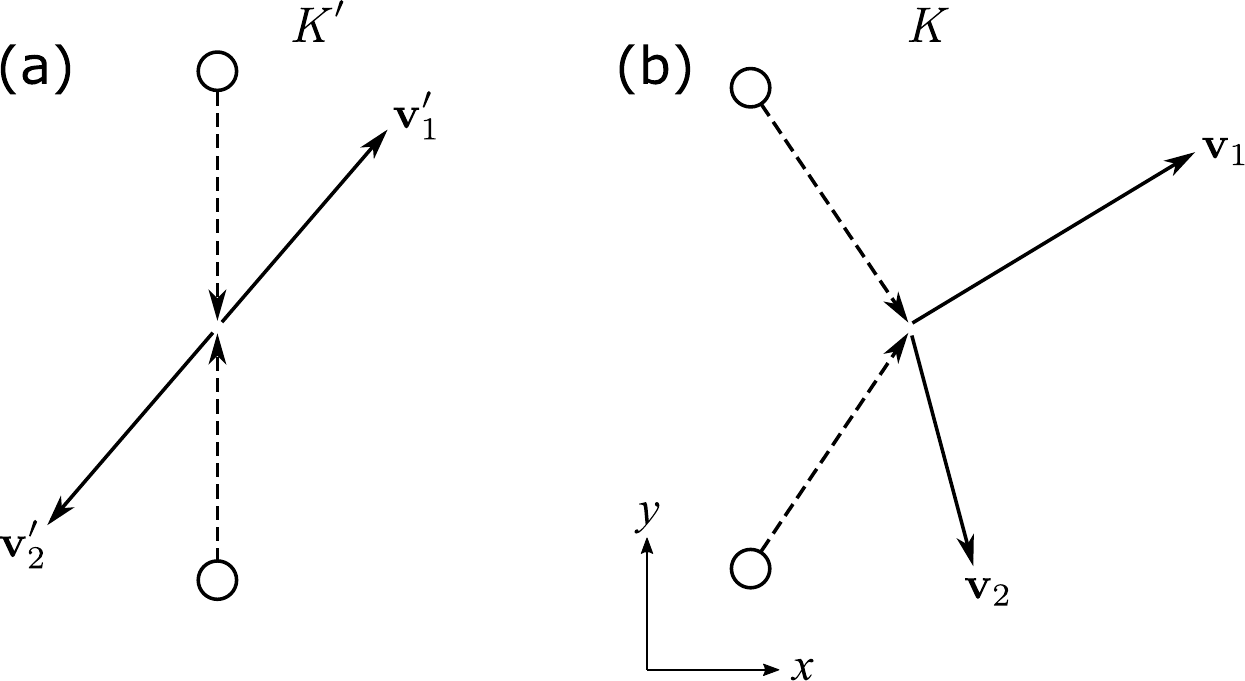}
\caption{Thought experiment adopted from the paper by
  Finkler~\cite{Finkler1996} that proves the relativistic
  formula for the momentum,
  eq.~(\ref{eq:prelativisticfinal}).\label{fig:Finkler}}
\end{figure}

Finkler~\cite{Finkler1996} considers
two identical particles that move
in the $xy$-plane with opposite velocities $\vecv'_1$
and $\vecv'_2 = -\vecv'_1$ relative to the center-of-mass frame $K'$,
see Fig.~\ref{fig:Finkler}a.
In an `unprimed' frame $K$, see Fig.~\ref{fig:Finkler}b, that
moves along the $x$-axis with respect to $K'$,
the $y$-component of the total momentum $p_{1y} + p_{2y}$
is equal to zero. Indeed, one may imagine that
the particles originally move along the $y$-axis in frame $K'$,
as shown by dashed lines,
and were scattered elastically. Then, in frame $K$,
the $y$-components of their momenta before the collision compensated
each other by symmetry, as seen in Fig.~\ref{fig:Finkler}b.
We write the $y$-component of a particle's momentum as
$p_y = p(v)\,v_y/v$, and express the vanishing of the total $y$-momentum
in frame $K$ as
\begin{equation}
  p(v_1)\,\frac{v_{1y}}{v_1} = - p(v_2)\,\frac{v_{2y}}{v_2} \, .
  \label{eq:Finkler1}
\end{equation}
The $y$-components of the particle velocities $\vecv_1$
and $\vecv_2$ in the `unprimed' frame $K$ can be transformed
to frame $K'$ using the rule~(\ref{eq:vy-vz-rule}),
\begin{equation}
  v_{1y} = v'_{1y} \, \frac{\gamma(v'_1)}{\gamma(v_1)} \, , \quad
  v_{2y} = v'_{2y} \, \frac{\gamma(v'_2)}{\gamma(v_2)} \, .
  \label{eq:Finkler2}
\end{equation}
Since $v'_1 = v'_2$ and $v'_{1y} = - v'_{2y}$ it follows
from eq.~(\ref{eq:Finkler2}) that
\begin{equation}
  \gamma(v_1)\,v_{1y} = - \gamma(v_2)\,v_{2y} \; .
  \label{eq:Finkler3}
\end{equation}
Finally, dividing each side of eq.~(\ref{eq:Finkler1})
by the corresponding side of eq.~(\ref{eq:Finkler3}) we find
\begin{equation}
  \frac{p(v_1)}{\gamma(v_1)\,v_1} = \frac{p(v_2)}{\gamma(v_2)\,v_2} \; .
  \label{eq:Finkler4}
\end{equation}
Varying the conditions of this thought experiment with
velocities and inclination angles in frame $K'$,
and the velocity of frame $K$ relative to $K'$,
we can independently vary $v_1$ and $v_2$ in eq.~(\ref{eq:Finkler4}).
Therefore, eq.~(\ref{eq:Finkler4}) implies that both sides are constants,
\begin{equation}
  F = \frac{p(v)}{\gamma(v)\,v}
  \label{eq:Finkler5}
\end{equation}
is independent of~$v$.
In other words, it is proven that the relativistic momentum $p(v)$
is proportional to $\gamma(v)v$. Using the non-relativistic limit
it is seen that $F\equiv m$ is the particle mass.

\subsection{Relativistic kinetic energy}
\label{subsec:relkinexamples}

In his unpublished lectures in the 1920s, Langevin uses
the collision scheme as in Fig.~\ref{Fig:bothframes}
to derive the relativistic formula for the kinetic energy.
Later, his arguments were reconstructed by Penrose, Rindler,
and Ehlers~\cite{Penrose1965, Ehlers1965}. Here, we only provide
a brief summary of their arguments.

\begin{figure}[ht]
\includegraphics[width=0.4\linewidth]{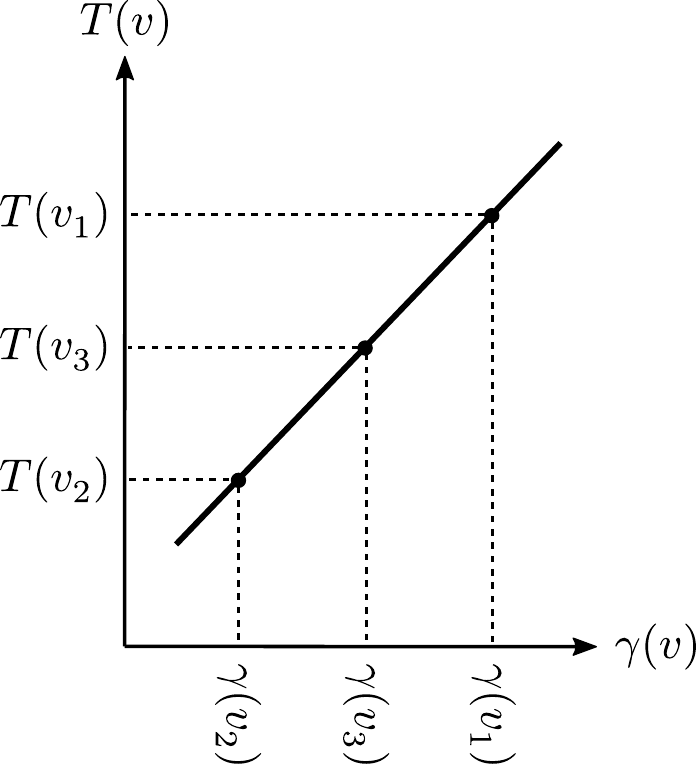}
\caption{Kinetic energy as a function of the
  $\gamma$-factor.\label{fig:Penrose}}
\end{figure}

Energy conservation during the elastic collision
in the 'unprimed' frame, see the right part of Fig.~\ref{Fig:bothframes},
states that $T(v_1) + T(v_2) = T(v_3) + T(v_4)$
where $T(v)$ denotes the kinetic energy.
Since $v_3 = v_4$ by symmetry, we have
\begin{equation}
  T(v_3) = \frac{T(v_1) + T(v_2)}{2} \; .
  \label{eq:Penrose1}
\end{equation}
Then, applying the rule~(\ref{eq:gamma-rule})
to the velocities $\vecv_1$, $\vecv_2$, $\vecv_3$, and $\vecv_4$,
one obtains
\begin{eqnarray}
  \gamma(v_1) &=& \gamma(v) \gamma(u) \left(1 + \frac{vu}{c^2}\right) \; ,
  \label{eq:Penrose2}\\
  \gamma(v_2) &=& \gamma(v)  \gamma(u)  \left(1 - \frac{vu}{c^2}\right) \; ,
  \label{eq:Penrose3}\\
  \gamma(v_3) &=& \gamma(v_4) = \gamma(v) \gamma(u) \; ,
  \label{eq:Penrose4}
\end{eqnarray}
whence
\begin{equation}
  \gamma(v_3) = \frac{\gamma(v_1) + \gamma(v_2)}{2} \; .
  \label{eq:Penrose5}
\end{equation}
When we plot the kinetic energy $T(v)$ versus $\gamma(v)$,
see Fig.~\ref{fig:Penrose}, eqs.~(\ref{eq:Penrose1}) and~(\ref{eq:Penrose5})
imply that the point $\bigl(\gamma(v_3),T(v_3)\bigr)$
in the plot
is always located exactly in the middle between the points
$\bigl(\gamma(v_1),T(v_1)\bigr)$ and $\bigl(\gamma(v_2),T(v_2)\bigr)$.
Since $v_1$ and $v_2$ can be varied independently,
the only possible shape of the curve $T(v)$ as a function of $\gamma(v)$
is a straight line,
\begin{equation}
  T(v) = c_1 \, \gamma(v) + c_2 \; .
  \label{eq:Penrose6}
\end{equation}
Here, $c_1$ and $c_2$ are some coefficients to be found from comparison
with the non-relativistic expression $T(v\ll c) \approx mv^2/2$,
namely, $c_1=mc^2=-c_2$.
Using these coefficients, eq.~(\ref{eq:Penrose6}) reduces to
the relativistic expression $T(v) = mc^2(\gamma(v)-1)$ for the
relativistic kinetic energy, see eq.~(\ref{eq:solutionTofv}).

\subsection{Energy-mass relation}

With relativistic formulas for the momentum and the kinetic energy
in our hands, we can go ahead and find the relation between mass and energy.
In the {\sl Feynman Lectures on Physics}~\cite{Feynman-Lectures-1}
this is done in the following way.

\begin{figure}[ht]
\includegraphics[width=0.6\linewidth]{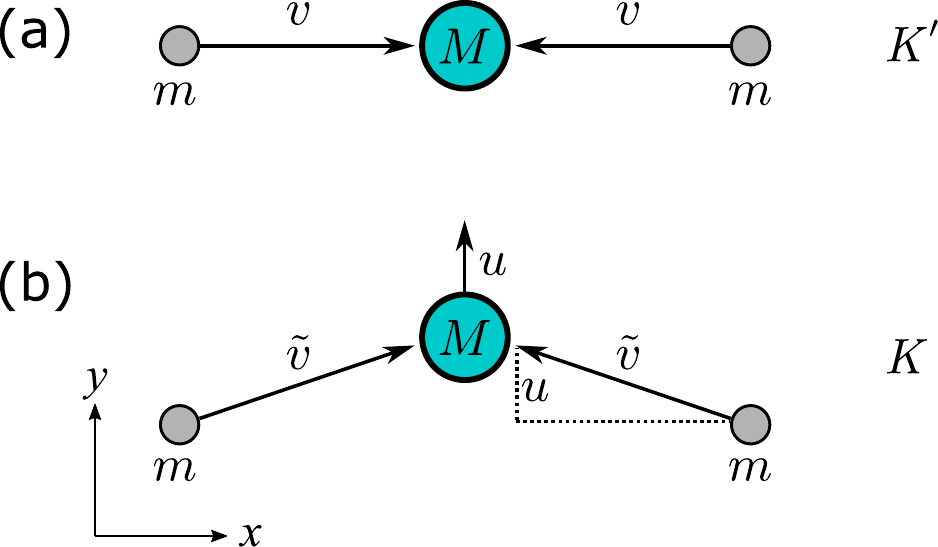}
  \caption{Thought experiment considered in the {\sl Feynman Lectures
      on Physics}~\cite{Feynman-Lectures-1}
    for obtaining the energy-mass relation.\label{fig:Feynman}}
\end{figure}

Imaging that two equal masses $m$ approach each other along the $x$-axis
with equal speeds $|\vecv|=v$. When they meet, they coalesce
into a larger body of some mass $M$, see Fig.~\ref{fig:Feynman}a .
Let us look at this process from another reference frame $K$,
see Fig.~\ref{fig:Feynman}b, that moves along the $y$-axis
relative to the center-of-mass frame $K'$ with a very small,
non-relativistic speed~$u$.
Conservation of the $y$-component of the momentum in the frame $K$ reads
\begin{equation}
  2 \gamma(\tilde v) m u = M u \; ,
  \label{eq:Feynman1}
\end{equation}
where $\tilde{v}$ is the speed of the initial masses in the frame $K$.
Neglecting the difference between $v$ and $\tilde{v}$,
we find the fused mass $M$ from eq.~(\ref{eq:Feynman1}),
\begin{equation}
  M = 2 \gamma(v) m \; .
  \label{eq:Feynman2}
\end{equation}
Therefore, the fused mass exceeds the original particle masses by
\begin{equation}
  \Delta m = M - 2m = 2m(\gamma(v)-1) \; .
  \label{eq:Feynman3}
\end{equation}
At the same time, the sum of the internal energies of the particles
have increased by
\begin{equation}
  \Delta E_0 = 2mc^2(\gamma(v)-1)
  \label{eq:Feynman4}
\end{equation}
because all the kinetic energy $T(v)=2mc^2(\gamma(v)-1)$
of the two masses, as seen from the center-of-mass frame $K'$,
transformed into internal energy. From comparison between
equations~(\ref{eq:Feynman3}) and~(\ref{eq:Feynman4})
we come to the conclusion that
\begin{equation}
  \Delta E_0 = \Delta m c^2 \; ,
  \label{eq:Feynman5}
\end{equation}
the mass defect formula~(\ref{eq:massdefect}).

\section{Conclusions}
\label{Sec:conclusions}

In this work we used relativistic kinematics of point
particles and two-particle collisions to derive the
energy-mass relation that reduces to Einstein's famous
formula~(\ref{eq:emcsquarebare}) for a particle at rest.
Our derivation offers several advantages over other derivations
briefly discussed in Sec.~\ref{sec:whatdidtheothersdo}.
\begin{enumerate}
\item Since it does not appeal to electrodynamics, it is conceptually simpler
  than Einstein's original argument that involves electromagnetic radiation.
  Moreover, using the relativistic Doppler formula,
  we can derive the energy of photons as massless particles and
  Planck's formula~(\ref{eq:almostdoneEhbaromega}) from an analysis of
  Compton scattering or from particle annihilation into two photons.
\item We address very simple geometries for the
  two-particle scattering, see Fig.~\ref{Fig:bothframes},
  which already provides us with the relativistic expressions for
  the relativistic expressions for particle momentum and its kinetic energy.
\item The mass defect formula straightforwardly follows from the
  decay of a massive particle at rest into two identical particle,
  see Fig.~\ref{fig:fission}. With the plain assumption
  that the internal energy of a point particle can only depend on
  its mass, Einstein's formula~(\ref{eq:emcsquarebare}) readily follows.
\end{enumerate}
The price of the conceptual simplicity is the use of
some standard elements of calculus,
namely (second-order) Taylor expansion and ordinary first-order
differential equations to find the unique solutions
of eqs.~(\ref{eq:definingeqforpofv}), (\ref{eq:solvethiseqforT}),
(\ref{eq:compton-energy-conservation}),
and~(\ref{eq:compton-momentum-conservation}).
The proof that the given expressions solve these equations requires
only elementary mathematics.

Students who are scared off by calculus and prefer
physically motivated shortcuts may resort to the literature, some of which
we briefly reviewed above. We presume that
the most straightforward `derivation' of Einstein's energy formula
starts from the four-vector of the relativistic velocity.
Scattering thought experiments show that
the three spatial components are conserved in a scattering experiment, i.e.,
they must be the particle momentum up to a mass factor.
Therefore, the `Babylonian approach' inspires
the notion that the zero component must be the particle energy,
up to a mass factor~\cite{Minkowski1910,Einstein1935}.

In this work we argue that the pedestrian but still Euclidean way
to Einstein's formula is neither short nor simple. Instead, it requires
the detailed analysis of scattering experiments and
some calculus to extract the correct formulas for the relativistic
momentum and kinetic energy.

\appendix

\section{Derivation of the momentum modulus}
\label{app:givepofv}

In this appendix we derive eq.~(\ref{eq:pofvwithconstant}).
We expand eq.~(\ref{eq:definingeqforpofv}) in a Taylor series in $u$
around $u=0$ up to first order in $u$. First, we find
\begin{equation}
p\left( \frac{v+u}{1+vu/c^2} \right) = p(v) + (1-v^2/c^2) p'(v) u + \ldots \; ,
\end{equation}
where the ellipsis denotes further terms that are proportional to $u^2$,
$u^3$, and so on. The prime denotes the derivative with respect to~$v$.
Next,
\begin{equation}
p\left( \frac{v-u}{1-vu/c^2} \right) = p(v) - (1-v^2/c^2) p'(v) u + \ldots
\end{equation}
and
\begin{multline}
  2 \frac{u}{\sqrt{v^2+u^2-v^2u^2/c^2}}
  p\left( \sqrt{v^2+u^2-v^2u^2/c^2} \right) = \\
  = \frac{2u}{v} p(v) + \ldots \; .
\end{multline}
We collect the terms to first order in $u$
in eq.~(\ref{eq:definingeqforpofv})
and find the condition
\begin{equation}
2 (1-v^2/c^2) p'(v) = \frac{2}{v} p(v) \; .
\end{equation}
Writing $p'(v)={\rm d}p/{\rm d}v$, this differential equation can be solved by
separation of variables,
\begin{equation}
\frac{dp}{p} = \frac{dv}{v(1-v^2/c^2)} \; ,
\end{equation}
so that the integration of both sides leads to
\begin{equation}
\ln p(v) = \ln \left[\frac{v}{\sqrt{1-v^2/c^2}}\right] + {\rm const} \; ,
\end{equation}
or
\begin{equation}
p(v) = C_p \frac{v}{\sqrt{1-v^2/c^2}}  \; ,
\end{equation}
which proves eq.~(\ref{eq:pofvwithconstant}).

\section{Derivation of the kinetic energy}
\label{app:prooftofv}

Eq.~(\ref{eq:solvethiseqforT})
defines the yet unknown function $T(v)$
for the kinetic energy.
As in appendix~\ref{app:givepofv},
we expand each term of this equation in a Taylor series around $u = 0$
to second order in~$u$,
\begin{eqnarray}
T\left(\frac{v+u}{1+vu/c^2}\right)
\approx  T(v) + u \left(1-\frac{v^2}{c^2}\right) T'(v) \nonumber \\
 + \frac{u^2}{2} \left(1-\frac{v^2}{c^2}\right) \left[ -\frac{2v}{c^2} T'(v)
  + \left(1-\frac{v^2}{c^2}\right) T''(v) \right] \; ,\\
T\left(\frac{v-u}{1-vu/c^2}\right)
\approx  T(v) - u \left(1-\frac{v^2}{c^2}\right) T'(v) \nonumber \\
 + \frac{u^2}{2} \left(1-\frac{v^2}{c^2}\right) \left[ -\frac{2v}{c^2} T'(v)
  + \left(1-\frac{v^2}{c^2}\right) T''(v) \right] \; ,\\
T\left(\sqrt{v^2 + u^2 - \frac{v^2u^2}{c^2}}\right)
\approx   T(v) + \frac{u^2}{2}
\left(1-\frac{v^2}{c^2}\right) \frac{T'(v)}{v} \; ,
\end{eqnarray}
where $T'(v)$ and $T''(v)$ are first and second derivatives
of function $T(v)$, and higher orders in the Taylor expansion were
ignored.
When inserted into eq.~(\ref{eq:solvethiseqforT}),
the constant and linear terms drop out, and
the quadratic terms lead to the differential equation
\begin{equation}
  -\frac{2v}{c^2} T'(v)
  + \left(1-\frac{v^2}{c^2}\right) T''(v) = \frac{T'(v)}{v} \; .
\end{equation}
We temporarily denote $T'(v)$ as $f$, and $T''(v)$ as $df/dv$
and find the first-order differential equation
\begin{equation}
  -\frac{2v}{c^2} f + \left(1-\frac{v^2}{c^2}\right) \frac{df}{dv}
  = \frac{f}{v}
\end{equation}
that is solved by separation of variables. It has the solution
\begin{equation}
f = f_0 \frac{v}{(1-v^2/c^2)^{3/2}}\; ,
\end{equation}
where $f_0$ is an arbitrary constant.
We recall that $f = T'(v)$, and integrate once more to find
\begin{multline}
  T(v) = \int^v {\rm d} x \, T'(x) = \int^v {\rm d} x\, f_0
  \frac{x}{(1-x^2/c^2)^{3/2}} = \\
=C_1 \frac{1}{\sqrt{1-v^2/c^2}}+ C_2 \; ,
\end{multline}
or
\begin{equation}
T(v) = \gamma C_1 + C_2\; ,
\end{equation}
where $C_1$ and $C_2$ are some constants.

We can fix $C_2$ by considering the particle at rest,
when $v=0$, $T=0$, and $\gamma=1$,
\begin{equation}
  0 = C_1 + C_2\; ,
\end{equation}
hence $C_2 = -C_1$ and
\begin{equation}
T(v) = (\gamma-1) C_1 \; .
\end{equation}
To determine $C_1$, we consider small but non-zero velocities $v/c \ll 1$.
The expansion of the $\gamma$-factor near $v= 0$ gives
$\gamma(v) \approx 1 + v^2/(2c^2)$ so that
\begin{equation}
T(v\ll c) \approx  \frac{C_1}{2c^2} \, v^2 \; .
\end{equation}
The non-relativistic formula reads $T^{\rm nr}= m v^2/2$,
see eq.~(\ref{eq:TandEpointparticleNR}),
so that we must set $C_1 = mc^2$. Hence,
\begin{equation}
T(v) = (\gamma-1) mc^2 \; ,
\end{equation}
as given in eq.~(\ref{eq:solutionTofv}).

\section{Derivation of photon energy and momentum}
\label{app:compton}

For better readability, we set the speed of light $c$ to unity within
this appendix.
Hence, equations~(\ref{eq:compton-energy-conservation})
and~(\ref{eq:compton-momentum-conservation}) take the simpler forms
\begin{multline}
  \gamma\left(\frac{v+u}{1+vu}\right) m + f\left(\omega
  \sqrt{\frac{1-u}{1+u}}\right)
  = \\
  = \gamma\left(\frac{v-u}{1-vu}\right) m + f\left(\omega
  \sqrt{\frac{1+u}{1-u}}\right) \; ,
  \label{eq:compton-energy-conservation-no-c}
\end{multline}
and
\begin{multline}
  \gamma\left(\frac{v+u}{1+vu}\right) m\,\frac{v+u}{1+vu}
  - g\left(\omega \sqrt{\frac{1-u}{1+u}}\right) =\\
  = -\gamma\left(\frac{v-u}{1-vu}\right) m\,\frac{v-u}{1-vu}
  + g\left(\omega \sqrt{\frac{1+u}{1-u}}\right) \; ,
 \label{eq:compton-momentum-conservation-no-c}
\end{multline}
where $\gamma(v) = (1-v^2)^{-1/2}$.

The Taylor expansions in
eq.~(\ref{eq:compton-energy-conservation-no-c}) at $u=0$ up to and including
linear terms in $u$ require
\begin{eqnarray}
  \gamma\left(\frac{v\pm u}{1\pm vu}\right) m
  &=& \gamma(v)m \left[ 1 \pm uv + {\mathcal O}(u^2) \right] \; ,\\
  f\left(\omega \sqrt{\frac{1\pm u}{1\mp u}}\right) &=&
  f(\omega) \pm u\omega f'(\omega) + {\mathcal O}(u^2) \; .
\end{eqnarray}
Substituting these expansions
into eq.~(\ref{eq:compton-energy-conservation-no-c}),
we see that the constant terms cancel each other,
and the linear terms lead to
\begin{equation}
  \gamma(v) mv =  \omega f'(\omega) \; .
  \label{appeq:fprime}
\end{equation}
The left-hand side of eq.~(\ref{appeq:fprime})
can be replaced with $g(\omega)$, see eq.~(\ref{eq:compton-v-omega}).
Hence, from eqs.~(\ref{eq:compton-energy-conservation-no-c})
and~(\ref{eq:compton-v-omega}) we obtain a differential relation
between the photon energy~$f(\omega)$ and its momentum $g(\omega)$,
\begin{equation} \label{eq:f-via-g}
\omega f'(\omega) = g(\omega) \; .
\end{equation}

Note that, for a massive particle traveling with some velocity~$v$,
the momentum-to-energy ratio fulfills $p/E=\gamma mv/\gamma mc^2 = v/c^2$.
Into the latter relation we might insert $v=c$
for a photon traveling with velocity~$c$.
Then, $g(\omega)/f(\omega) \equiv p/E = 1/c$ immediately follows, i.e.,
$g(\omega) = f(\omega)$ when $c=1$.
Therefore, eq.~(\ref{eq:f-via-g}) takes the form $\omega f'(\omega) = f(\omega)$
that immediately gives rise to the conclusion that the photon energy $f(\omega)$
is proportional to the frequency $\omega$.
However, we shall not follow this shortcut. Instead, we will
employ the momentum conservation law,
eq.~(\ref{eq:compton-momentum-conservation-no-c}).

The Taylor expansion of the terms contributing to
eq.~(\ref{eq:compton-momentum-conservation-no-c}) at $u=0$ require
\begin{equation}
  \gamma\left(\frac{v\pm u}{1\pm vu}\right) m\,\frac{v\pm u}{1\pm vu}
  = \gamma(v)m \left[ v \pm u + \frac{u^2v}{2} + {\mathcal O}(u^3) \right]
\end{equation}
and
\begin{multline}
  g\left(\omega \sqrt{\frac{1\pm u}{1\mp u}}\right) =\\
  = g(\omega) \pm u\omega g'(\omega) + \frac{u^2}{2} \left[ \omega g'(\omega)
    + \omega^2 g''(\omega) \right] + {\mathcal O}(u^3)
\end{multline}
up to and including second order in~$u$.
Substituting these expansions into
eq.~(\ref{eq:compton-momentum-conservation-no-c}), we see that the constant
terms reproduce the known eq.~(\ref{eq:compton-v-omega})
whereas the terms proportional to $u$ cancel each other.
Collecting the terms proportional to $u^2$ we find
\begin{equation}
  \gamma(v)mv = \omega g'(\omega) + \omega^2 g''(\omega) \; .
  \label{appeq:gprimeeq}
\end{equation}
In eq.~(\ref{appeq:gprimeeq}), the left-hand side
can be replaced by $g(\omega)$
due to eq.~(\ref{eq:compton-v-omega}). Therefore,
we arrive at a differential equation for the function $g(\omega)$,
\begin{equation} \label{eq:g-equation}
\omega^2 g''(\omega) + \omega g'(\omega) - g(\omega) = 0  \; .
\end{equation}
This is a homogeneous second-order linear equation
with the two independent solutions
$g_1(\omega) = \omega$ and $g_2(\omega) = \omega^{-1}$.
Hence, the general solution is
\begin{equation} \label{eq:g-solution}
g(\omega) = C_1 \omega + \frac{C_2}{\omega} \; ,
\end{equation}
where $C_1$ and $C_2$ are arbitrary constants.
Substituting this solution into eq.~(\ref{eq:f-via-g})
and integrating over~$\omega$
leads to the function $f(\omega)$,
\begin{equation} \label{eq:f-solution}
f(\omega) = C_1 \omega - \frac{C_2}{\omega} + C_3 \; ,
\end{equation}
where $C_3$ is yet another constant.
When we return to physical units, we have to divide the right-hand side of
eq.~(\ref{eq:g-solution}) by $c$ to account the difference
between the unit of energy
and that of momentum. In this way, we obtain eqs.~(\ref{eq:f-omega-via-C})
and~(\ref{eq:g-omega-via-C}) of the main text.

\bibliography{emcbib}

\begin{thebibliography}{23}
\bibitem [{Bab()}]{Babel}%
  \BibitemOpen
  \href@noop {} {}\bibinfo {note} {The Babylonians knew $\sqrt{2}\approx
  1.41421$ with six-digit accuracy but it is unknown how they came to this
  result. Naturally, this guess can be tested by calculating
  $1.41421^2=1.9999899241$. Euclid showed that $\sqrt{2}$ is irrational, and
  the Heron and Bakhshali algorithms permit to compute it numerically very
  efficiently.}\BibitemShut {Stop}%
\end{thebibliography}

\end{document}